# High accuracy Spin Hall Effect Induced Spin Accumulation detection in MOKE Measurements

Emanuele Longo[1,*], Josep Fontcuberta[1], Paolo Vavassori[2,3,**]

1. Institut de Ciència de Materials de Barcelona (ICMAB-CSIC), Campus UAB, Bellaterra, Catalonia 08193, Spain
2. CIC nanoGUNE BRTA, E-20018 Donostia-San Sebastian, Spain
3. IKERBASQUE, Basque Foundation for Science, E-48009 Bilbao, Spain

*elongo@icmab.es, **p.vavassori@nanogune.eu

**Abstract**

Charge to spin (orbital) momentum conversion phenomena enclose great potential for advancing applications in spin/orbitronics. Although current-induced magnetic moment accumulation is crucial both for fundamental understanding and practical applications, direct quantifications are scarce. Optical polarization measurements, namely magnetooptical Kerr rotation (MOKE) ($\theta_K$), have been used to get direct evidence of magnetic accumulation perpendicular to a current flow density (J) in late transition meals (Pt) as well as in light transition elements (Ti, Cr) and used to conclude evidence of spin or orbital moment accumulation. However, discrepancies of the reported $\theta_K/J$ values, exceeding one order of magnitude, together with early claims that conventional MOKE experiments were not a suitable tool, is prompting revisions of methods and results. Here, we report on a new methodology for MOKE measurements that solves known bottlenecks. We obtain a sensitivity of $(354 \pm 27)$ nV/nrad and use the designed protocol to measure $|\theta_K^S/J| = (7.92 \pm 1.94)$ nrad/($10^7$ A/cm$^2$) and $|\theta_K^P/J| = (6.89 \pm 1.74)$ nrad/($10^7$ A/cm$^2$) in a 50 nm thick Pt bar for S and P polarized incident light, respectively. The extracted value of $|\theta_K^S/J|$ is significantly smaller, about a 7-fold reduction, than previous results on a nominally identical device. Given that differences in the microstructure of Pt films cannot account for a such large discrepancy, this implies that experimental procedures and models should be revised accordingly.

**Introduction**

Over the past several decades, the Magneto-Optic Kerr Effect (MOKE) has emerged as a powerful and widely adopted technique for probing magnetic phenomena in condensed matter systems. Traditionally, MOKE has been employed to study ferromagnetic materials (e.g., Ni, Fe), where large Kerr rotation angles ($\theta_K$) in the microradian - milliradian range can be measured thanks to their robust spin polarization and spin-orbit coupling (SOC). [1,2] In recent years, research has further advanced the frontier of spintronics and orbitronics by attempting to directly detect the current-induced accumulation of magnetic moments in non-magnetic materials (NM), whether spin or orbital. While spintronics has traditionally focused on the manipulation of electron spin, orbitronics aims to harness the often-overlooked orbital angular momentum of the electron, offering an additional degree of freedom for next-generation information technologies. [3,4]

Spin Hall Effect (SHE) and Orbital Hall Effect (OHE) are responsible for the conversion of a charge current into a transverse spin or orbital currents, constituting probably the most technologically relevant phenomena in spin-orbitronics. [3–7] Thanks to its versatility, MOKE become a fundamental tool to investigate SHE and OHE phenomena in single layer thin films, simplifying the detection of magnetic moments accumulation. Indeed, more conventional methods, such as spin-torque and spin-pumping usually require combination with other materials (i.e., introduction of interfaces) and/or non-local detection (i.e., spin-orbit coupling, proximity effect, scattering effects, etc.), increasing the complexity of the data interpretation.

Indeed, MOKE was successfully used long ago to detect current-induced spin accumulation in semiconductors, such as AsGa and InGaAs. [8] However, as current-induced local spin density in metals is much smaller due to the shorter spin relaxation time, its magnetic moment accumulation fingerprints are weaker, thus challenging experiments. For these reasons may not be surprising that earlier claims of observation of current induced spin accumulation in metals [9] were much disputed, [10,11] until Stamm et al. [12] reported more convincing MOKE observation of current-induce spin accumulation in the Pt and W heavy metals, and Ortiz *et al*. [13] in Au. A relevant example can be found in the manuscript of Stamm *et al*. [12], where the magnitude of the SHE has been investigated for Pt and W single layers with different thicknesses, monitoring the Kerr rotation induced by the system under the application of different charge current density (J). More recently, also light materials with low-SOC such as Ti and Cr have been reported to induce a detectable Kerr rotation when a charge current is flowing throw them. [14,15] Such a magneto-optic effect was claimed to originate due to the accumulation of orbital magnetic moments at the surfaces of thin films due to OHE. However, the absolute Kerr rotation measured in NM remains extremely small, typically in the nanoradian range for J of the order of $10^7$ A/cm$^2$, since any induced magnetization arising from spin accumulation is orders of magnitude weaker than the intrinsic magnetization in ferromagnetic materials. The reported $\theta_K$ values for nearly identical Pt thickness differ by more than an order of magnitude (2 - 50 nrad/($10^7$A/cm$^2$)) [12,14,16], which may illustrate the extremely demanding nature of these measurements. The findings expressed above underscore a critical challenge: while spin and orbital accumulation can be in principle directly detected by MOKE, their magneto-optical signature is faint when measured in NM. This poses a major challenge in discerning genuine Kerr signals from experimental artifacts, such as noise of voltage/current biases, that can lead to unwanted thermal effects, thermoreflectivity fluctuations, or minute geometrical misalignments in the optics leading to systematic errors, that may dramatically impact the accuracy of the measurements. Indeed, the latter could result in signals displaying the same symmetry as the sought tiny MOKE effect upon modifying the experiment conditions (e.g., reversing the incidence and reflected path, etc.) as done in the experiments reported in literature, that can mimic spin/orbital magneto-optic signals, leading to the over-/under-estimation of physical effects or even to erroneous interpretations.

To address these challenges, we devised a new magneto-optical measurement protocol for the detection of SHE specifically designed to minimize the risk of artifact-induced misinterpretation. Our approach incorporates redundant experimental crosschecks and rigorous controls for spurious effects arising from thermal fluctuations and optical misalignments. By enforcing strict symmetry compliant validation criteria, we ensure that the observed Kerr rotation signals can be attributed to genuine spin/orbital magnetic moment accumulation in a single layer system. In order to provide a useful and directly applicable experimental methodology, we conducted our investigation on single layer Pt thin film. A Kerr rotation value of $(6.89 \pm 1.74)$ nrad/($10^7$ A/cm$^2$) is measured for a 50 nm thick Pt thin film, constituting only about 13.8% of the ~ 50 nrad/($10^7$ A/cm$^2$) reported in Ref. [12] for the same material. In this work, Pt is employed as the charge-to-spin conversion material, serving as a benchmark in state-of-the-art MOKE experiments owing to its chemical and electronic robustness and reproducible behavior. Nevertheless, the proposed methodology is universally applicable to other materials, including both heavy and light metals, where magnetic moment accumulation may be of orbital and/or spin origin.

**Material Preparation and Experimental Setup**

To perform MOKE experiments, the Pt thin films are grown on thermally oxidized Si(001)/SiO$_2$ (100 nm) substrate by using DC magnetron sputtering at room temperature with 3 mTorr Ar. Out of the Pt continuous film, we prepared 50 x 40 μm rectangular stripes using photolithography and lift-off technique. X-ray diffraction and reflectivity are used to prove the crystalline nature of the Pt films, which are characterized by an out-of-plane texture along the [111] crystalline direction, an electronic density of 4.957 $e^-$/Å$^3$ (nominal 5.191 $e^-$/Å$^3$) and a roughness of 0.23 nm (Supp. Inf. Fig. S1). Atomic force microscopy images collected on different samples areas confirmed the continuous morphology and the roughness value of the Pt film, in accordance with X-ray results (Supp. Inf. Fig. S2). Resistivity (ρ) versus temperature measurements acquired in Hall-crossbar configuration are compatible with those already reported in literature for the same material, where ρ ~ 42 $\mu\Omega \cdot$ cm at room temperature (Supp. Inf. Fig. S3). [12,14,16] To perform MOKE measurement the Pt stripes are contacted on a custom board using Ag wires and a function generator is used to apply a sinusoidal current of frequency (ω), reaching a density $J = 1 \cdot 10^7$ A/cm$^2$. As customary, to enhance the detection sensitivity, the signal voltage detected by a low-noise avalanche photodiode, after the reflection reference signal subtraction, is locked at the frequency of the injected AC current using a lock-in amplifier.

The contacted Pt stripe is placed in an optical setup equipped with an ultralow noise red laser ($\lambda = 635$ nm) with unmatched low noise performance (typical RMS noise less than 0.06% for detector bandwidth from 10 Hz to 10 MHz) used to shine the sample surface with polarized light through a large numerical aperture polarization maintaining objective to focus the beam in micron-size an elliptical spot (~ 3 μm x 4 μm) at an incidence angle of 40º.

**Results and discussion**

In a conventional experiment for the detection of the current-induced magnetic moment accumulation in a NM metallic layer, a charge current is injected into a lithographed structure and a polarization-sensitive magneto-optical Kerr effect (MOKE) equipment is adopted to track down the tiny polarization change induced by SHE or OHE (spin and orbital Hall effect, respectively), named MOKE-HE. Since the signal is extremely small, a lock-in amplifier detection is used, where the amplitude of the AC injected current is modulated at a frequency ω (i.e., I(ω)) and the output signal subsequently demodulated at the same frequency. Such experiment requires a considerably high current density J, of the order of few $10^7 A/cm^2$, [12,14,16] which induces a temperature (T) increase of the material (∼ 10 − 20 K), and results in an optical signal variation (i.e., thermoreflectivity) producing a signal at 2ω. Moreover, a differential detection scheme is generally used in the attempt to remove any effect due to laser intensity instability and thermal drift (e.g., environment and current induced, or the potential presence of current biases). Finally, to single out the MOKE-HE signal from residual spurious effects, measurements are repeated with the same polarization conditions but inverting the light path as well as swapping the polarity of the AC current. Limitations of these approaches include three main aspects: a) the limited sensitivity and noise of the differential photodetectors, that impact the signal detection level; b) the difficulties of tracking a tiny magneto-optic signal oscillating at ω buried in a nearly three orders of magnitude larger signal [12] arising from thermoreflectance that, although it is oscillating mainly at 2ω, it also produces a leakage signal at ω (as well as at 3ω, although this latter is not relevant for the detection of the signal) that mixes with the signal of interest. As a consequence, large values of the time constant (TC) for the lock-in detection (e.g., TC =15 mins) [12] as well as taking many averages are required, making the measurements extremely time consuming (e.g., 24 - 48 h lasting measurements) [12] and drift-sensitive; c) the determination of a tiny light polarization rotation due to MOKE-HE in the film, implicitly assumes that the polarization state of the light impinging the sample and any depolarization along the light path are known and accounted for with extremely high accuracy.

To tackle the afore-mentioned experimental challenges, we developed a detection approach based on three steps, constituting a novel MOKE-HE measurement strategy, as described in the following.

**Step 1.** To solve the criticality a), we use a simple approach based on the use of a polarization analyzer positioned almost at the extinction angle (∼ ± 0.7° from extinction) and a nearly-photon counter detector (Avalanche Photodetector, APD). As shown in Ref. [17], in this optical configuration and for a depolarization factor of 2 x $10^{-4}$ close to that in our experiment (as discussed below in the manuscript), simple Jones matrix calculations [12,14–16] show that, neglecting the noise figure and optimum operation conditions of the photodetector, the maximum of the MOKE sensitivity is reached for the analyzer set to an angle of ∼ ± 0.7° from extinction. Detecting a light power (P ∼ 100 nW) at such a small angle would require an amplified photodiode operated at high gain, which would be too noisy. One solution to this problem is to use an APD, which is known to have the required gain but with a much lower spectral noise density (∼ 10 times less) than an amplified photodetector (see details in Supp. Inf. Fig. S4). To further improve the signal stability, we detect part (1/10) of the reflected signal intensity with a second conventional photodetector and use it as a reference to remove slow drifts, and we use an ultra-low noise (ULN) red laser, which has an unrivaled low noise stability. This experimental configuration guarantees an extremely stable signal and a large sensitivity to polarization changes (up to 354 mV/mrad in our setup, see below). The scheme of the experimental setup is sketched in Figure S4 in the Supporting Information, together with details on the used opto-electronic instrumentation.

**Step 2**. As a consequence of point b), the reflected light would introduce spurious $\omega$ contributions to the optical signal detection as due to thermoreflectance instabilities/fluctuations (e.g., consequence of T variation) and/or bias fluctuations of the current (or voltage) source. Such noise is commonly larger than the typical signals (expected to be in the order of 10 - 20 μV in our setup assuming a $\theta_K \sim 50$ nrad) [12,14,16] due to the magneto-optical MOKE-HE we have to identify. Indeed, according to MOKE-HE experiments reported in literature, a resolution down to 5 nrad is needed, that, with a sensitivity of ~ 350 mV/mrad = ~ 0.35 μV/nrad as in our experiment (a ten-fold improvement of the ~ 0.025 μV/nrad in the experiments previously reported [12]), makes it necessary to detect signals with at least ~ 1.5μV uncertainty. This condition can be reached by using large time constants (TC) at the lock-in detection as well as taking many averages, as commonly done. [12,14,16] However, the use of large TC makes the measurements extremely time consuming and, most importantly, drift-sensitive. Interestingly, the situation improves if a small bias is intentionally applied to I($\omega$) and swept from positive to negative values. We first observed that the application of a small dc-bias does not increase appreciably the noise arising from fluctuations (resulting in a lock-in output noise of ~10 μV in our experiments, with and without bias). In this case, the lock-in output is dominated by the bias induced signal at $\omega$, which varies linearly with the bias amplitude (see the detailed discussion below). Therefore, the plot of the acquired lock-in signals at different values of the bias must display a perfect linear dependence, which is not fulfilled if a drift has occurred during the measurements. This observation led us to the second step of our approach, constituted by the controlled sweeping of a dc-bias current added to I($\omega$), as described in the following.

In the simplest approach, the light intensity detected as a voltage $V^D$ by the APD, when a current I($\omega$) is injected through the sample using a dc-biased oscillating current (I($\omega$) = (v$_{bias}$ +v$_0$($\omega$)/r(T)), with r(T) the resistance of the device) can be written as: $V^D = V^O(\omega) + V^{th}(\omega, 2\omega)$. The first term $V^O(\omega)$ contains the signal generated by the APD and related to the polarization variation of the P-polarized or S-polarized ($\vec{E}$ field of incident radiation parallel or perpendicular to incidence plane, respectively) incident light after reflection from the sample and transmission through the analyzer, resulting from the genuine Kerr effect ($V^K(\omega)$). This term contains also contributions from any non-perfect polarization alignment (not pure P-/S-polarization), the finite polarization rejection rate of the polarizers and the unavoidable depolarization of the light beam due to optic elements ($V_{pm}$). Therefore: $V^O(\omega) = V^K(\omega) + V_{pm}(\omega)$. Assuming a "perfect" alignment of optical elements (see **Step 3** below), $V_{pm}(\omega)$ becomes negligibly small. The thermoreflectance term is $V^{th}(\omega, 2\omega) = R_{s,p}(T) [v_{bias} +v_0(\omega)]^2$, where $R_{s,p}(T)$ contains the polarization-dependent change of reflectance when the sample temperature rises under I($\omega$) flow. Therefore: $V^{th}(\omega, 2\omega) = R_{s,p}(T) (v^2_{bias}+ 2v_{bias}v_0(\omega)+v_0(2\omega))$. Whereas the last term is the common thermoflectance signal occurring at $2\omega$, which has a typical amplitude much larger than any magneto-optic effect collected by $V^{MO}(\omega)$ (see Supp. Inf. Fig. S5), the term $v_{bias}v_0(\omega)$ oscillates at the same frequency (and phase) as the magneto-optic signal.

Summing up:

$$V^D = V^{MO}(\omega) + V^{th}(\omega, 2\omega) = [V_{pm} + V^K(\omega)] + [R_{s,p}(T) (v^2_{bias}+ 2v_{bias}v_0(\omega)+v_0(2\omega))]$$

By using a synchronous and phase locked detection at $\omega$, we obtain

$$V^{lock-in}(\omega) = V^K(\omega) + 2 R_{s,p}(T) v_{bias}v_0(\omega)$$

We now introduce the experimental observation that $V^{th}(\omega, 2\omega)$ contains a fluctuation term $\delta V_{rand}(\omega-\delta\omega \leftrightarrow \omega+\delta\omega)$, which is mainly traceable to T fluctuations via r(T) and to the fact that $R_{s,p}(T)$ might not be a linear function. In a first approximation, sufficient for the present discussion, we can lump $\delta V_{rand}$ in a single term at $\omega$, $\pm|\langle 2\delta R_{s,p}\rangle_{rand}|v_0(\omega)$, where $|\langle 2\delta R_{s,p}\rangle_{rand}|$ is the maximum amplitude of an equivalent and randomly variable harmonic contribution, $\delta V^\omega_{rand}$, to $V^{lock-in}(\omega)$. Adding this term related to fluctuations and anharmonicities of the thermoreflectance, we obtain

$$V^{lock-in}(\omega) = V^K(\omega) + 2 R_{s,p}(T) v_{bias}v_0(\omega) \pm|\langle 2\delta R_{s,p}\rangle_{rand}|v_0(\omega)$$

If $v_{bias} = 0$, in the typical experiments (large current densities leading to temperature increase of > 10K), the lock-in signal is dominated by this fluctuating term (indeed, our measurements show that $|\langle 2\delta R_{s,p}\rangle_{rand}| v_0(\omega) \gg V^K(\omega)$, making clear that a reliable retrieval of $V^K(\omega)$ extremely challenging even using very large values of TC. In contrast, by applying a $v_{bias}$ sufficiently large, i.e., at least as large as $2R_{s,p}(T)v_{bias} > |\langle 2\delta R_{s,p}\rangle_{rand}|$, we observed that the fluctuating term $|\langle 2\delta R_{s,p}\rangle_{rand}|v_0(\omega)$ remains substantially the same, as mentioned above, thereby enabling the extraction of the signal $V^K(\omega) + 2 R_{s,p}(T)v_{bias}v_0(\omega)$ with a large signal to noise ratio. Finally, since $V^{lock-in}(\omega)$ is linear in $v_{bias}$, $V^K(\omega)$ can be retrieved with great precision through the extrapolation to $v_{bias}= 0$ of the linear fit parameters of a sequence of measurements of $V^{lock-in}(\omega)$ obtained by sweeping $v_{bias} \neq 0$ from + $v_{bias,max}$ to - $v_{bias,max}$. A detailed description of the effect of the voltage bias superposition on the injected AC current is reported in Supporting Information (Fig. S5). The occurrence of a significant drift during the measurements (i.e., leading to unreliable measurements), would produce a non-linear $V^{lock-in}(\omega)$ vs. $v_{bias}$ dependence, easily identifiable and that would lead to discard them. In addition, the slope of the $V^{lock-in}(\omega)$ vs. $v_{bias}$ linear dependence should marginally depend on the polarization selected for the measurements providing, therefore, a precious feedback about the precision of the optical alignment in the experiments (see next Step 3 and the Supporting Information, in particular Figs. S7 and S8).

**Step 3**. As indicated in point c) above, a very critical issue, is the polarization of the incident beam. The commonly adopted approach to discern MOKE-HE signals is to set a specific linear polarization of the incident beam, S or P, and detect any polarization and intensity change upon reflection (note that a polarization change is also measured as an intensity change) at oblique angle, and finally identifying the MOKE-HE signal based on its symmetry upon swapping the light path. The underlying idea is that MOKE-HE signal would change sign upon inverting the path, while any other contribution to the signal (e.g., thermoreflectivity) will not. The same would hold true by changing the incident polarization between P and S, and also, as it is done in our detection approach, by changing the analyzer angle from + $\phi_0$ to - $\phi_0$ from extinction. All this is true if the polarization of the incident beam is exactly P or S polarized, which means there is a perfect alignment, parallel for P and orthogonal for S polarization, between the electric field of the incident radiation and the reflection plane. Ensuring that these condition, that we name here on as "*polarization purity*", are met as close as possible in real experiments, it requires a special care and a specific alignment protocol, that has not been typically specified and used (i.e., Refs. [12,14,15]). Simply relying on the goniometer of a polarizer and checking the incidence plane orientation visually, or even with the aid of pinholes or diaphragms, is not enough when it comes to nano-radiant polarimetric metrology. At variance with the discussion about points a) and b), which is related to enhanced sensitivity and noise suppression, the lack of *polarization purity* will produce signals that will display the same symmetry as the targeted MOKE-HE signal upon changing the experimental conditions as described above. In other words, the lack of *polarization purity* is particularly treacherous as it gives rise to systematic spurious signals virtually indistinguishable from the MOKE-HE one and thus affect the accuracy of any $\theta_K$ derived.

The effect of a spurious polarization introduced by a tiny tilt between the incident polarization and the plane of incidence is thoroughly analyzed in the Section **7** of the Supporting Information. The key results of the analysis indicate that even a slightly tilted reflection plane (± 0.1°) results in a MOKE sensitivity, and thus a MOKE signal, that is markedly different for + $\phi_0$ and - $\phi_0$ (see Figs. S6a and S6c). Despite a common practice is to measure at either + $\phi_0$ or - $\phi_0$, this approach leads to the generation of a MOKE signal that overestimate or underestimate the real magnetic contribution, even in case of a minute optical misalignment (i.e., without ensuring a sufficient *polarization purity*).

We illustrate this effect by plotting in Fig. 1 what happens if only one detection angle $\phi_0$ is used and the light beam path or the current polarity are reversed. As sketched in Fig. 1(a), a + 0.1° tilt of the plane of incidence with respect to the polarization plane is applied, and in Fig. 1(b) the MOKE sensitivity $S(\phi)$ is depicted (see Supplementary Information S7 for definitions and derivation of the appropriate expressions).

It is evident from Fig. 1(b), that when using analyzers at ± $\phi_0$ angles from extinction, the MOKE signal differs and the difference is carried along upon swapping the light path or the current, representing a systematic error that produces the overestimation or underestimation of the real MOKE signal. For instance, in the case depicted in Fig. 1, the MOKE signals recorded at ± $\phi_0$, would differ by about 34%. This difference vanishes when any misalignment is suppressed (see Fig. S6b) and will increase to 80% with a tilt of + 0.2°. Note that + 0.2° is still a minute misalignment, that cannot be easily identifiable. Such an effect may not be introduced only by an imprecise setting of the polarizer's axes, but can arise from imperceptible tilts of optical elements or displacements of the beam with respect to the symmetry axes and planes of optical elements and could cause systematic signals showing the same symmetry and of the same entity, or even larger, as those of MOKE-HE. Therefore, a very precise alignment of the polarization and incidence plane is mandatory to ensure accurate measurements by suppressing systematic errors on $\theta_K$ when using measurement relying on light-path reversal as a mean to isolate a MOKE signal based on its symmetry (asymmetry upon reversing the path). As substantiated below, the method proposed here relaxes this constraint being self-correcting.

To ensure that the conditions of *polarization purity* are met as close as possible, we applied a recursive Malus law approach, which aims at removing any residual S-component when P polarization is meant, and residual P-component when S polarization is meant. Fig. 2a shows the schematic of the setup. We mounted the polarizer (Polarizer 1 in Fig. 2a, P1 in the text below) and the analyzer (Polarizer 2 in Fig. 2a, P2 in the text below) on a pair of high precision automated rotation stages (~ $5\times10^{-3}$ deg, see Section **5** of the Supporting Information). Initially, we set the polarization of the incident beam rotating the P1 parallel (P polarization) or vertical (S polarization) with respect to the horizontal plane of the experimental setup (the plane of the optical table), which ideally would correspond to the plane of incidence of the light (angle $\Phi_0^{P1}$ of the goniometer). The analyzer P2 is then set at extinction, $\Phi_0^{P2}$, finding the minimum APD signal $V^D(\phi)$ (Malus law, $V^D(\phi) \approx \sin^2\phi \approx \phi^2$) where now we define $\phi$ as the angle of the analyzer axis respect to extinction, namely $\phi = \Phi^{P2} - \Phi_0^{P2}$. Since, as we mentioned above, the real incidence plane of the light might not coincide with the horizontal plane, we recorded a Malus law by scanning the angle of the polarizer while keeping the analyzer at the extinction angle found initially. We then set the polarizer at the new angle of extinction $\Phi_1^{P1}$ and we record another Malus law by scanning the angle of the analyzer, to obtain a new extinction angle $\Phi_1^{P2}$. The sequence is repeated obtaining a sequence of $\Phi_i^{P1}$ and $\Phi_i^{P2}$ and the process stops when $\Phi_{i+1}^{P1} = \Phi_i^{P1}$ and $\Phi_{i+1}^{P2} = \Phi_i^{P2}$ within the experimental uncertainty. In these conditions, the minima of the Malus laws for both the polarizer and analyzer reach their lowest value and we ensured that achieved the condition of *polarization purity* as close as possible. The result of this recursive method is shown in Fig. 2b, which shows the Malus laws for "purified" (i.e., no spurious S or P polarization components, respectively) P (blue symbols) and S (black symbols) incident polarization. As an internal consistency check, one should find that the conditions for P and S polarization, $[\Phi_i^{P2}]_P$ and $[\Phi_i^{P2}]_S$ respectively, are achieved at 90º from each other ($[\Phi_i^{P2}]_P - [\Phi_i^{P2}]_S = (90.02° \pm 0.005)°$ in Fig. 2b). This is the first step for any set of measurements at a given incident polarization. The small difference between the minima of the plots in Fig. 2b is due to the slightly lower reflectivity at 40º of incidence for P polarization as compared to S polarization (Fresnel equations). This alignment is performed running a current density $J(\omega) = 10^7$ A/cm$^2$ through the device to make sure that there is no change in the *polarization purity* occurring due to the current and/or the bias. To convert the measured signals in a Kerr rotation angle $\theta_K$, we calculated the derivative dSig ($\phi$)/d$\phi$ of the Malus laws in Fig. 2b. Defining the new angle $\phi$ with respect to extinction (see Fig. 2c), the plots of dSig ($\phi$)/d$\phi$ are shown in Fig. 2d and allow us to calibrate the setup. For the analyzer angles $\pm \phi_0 = \pm 0.7°$ used in our experiments (shown if Fig. 2d) we obtain a conversion factor of $(2.04 \pm 0.01)$ V/° = $(118 \pm 9)$ mV/mrad. Since we used a differential amplifier with a gain G = 3 to subtract a reference signal of an amplified photodiode from the avalanche photodetector signal (see Fig. S4 of the SI for details), the final sensitivity of the setup is $([3 \times 118] \pm 9)$ mV/mrad = $(354 \pm 27)$ nV/nrad.

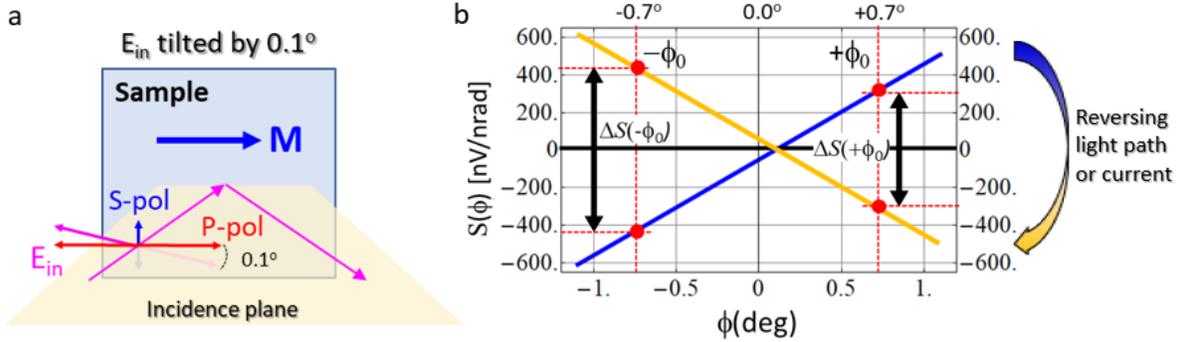

**FIG. 1.** (a) Sketch of the incidence plane of the light (shadowed yellow area) tilted by $\phi = + 0.1°$, with respect to the polarization plane. (b) $\phi$-dependent MOKE sensitivity $S(\phi)$ for two reversed directions of the light path (blue and orange lines). In magenta and black the corresponding $\Delta S(\pm\phi_0)$ obtained reversing the light path or current, for incidence angles from extinction at $\phi_0$= -0.7° and +0.7° respectively, are shown. The calculations indicate that substantially different values of the MOKE sensitivity $S(\phi)$, and thus $\Delta S(\pm\phi_0)$, are obtained depending on the sign of the sign of the detection angle $\phi_0$ (about 34% difference).

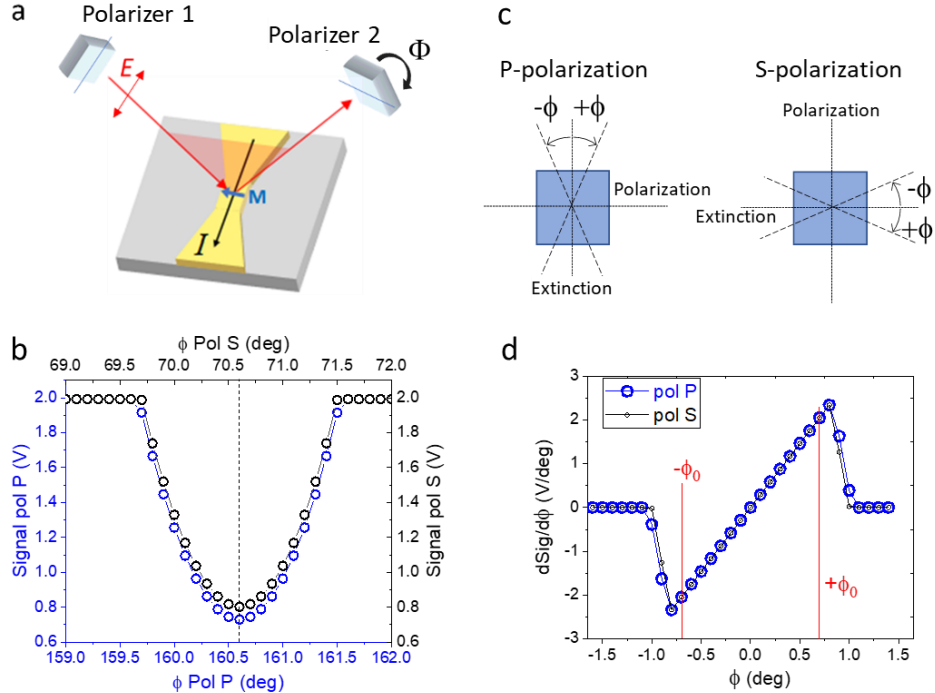

**FIG. 2.** (a) Sketch of the MOKE configuration scheme. The polarized light with an electric field E oscillating along a specific direction set by Polarizer 1 impinges on the sample surface. The reflected light is collected by Polarizer 2 (analyzer). The black arrow labeled by the letter "I" indicates the charge current applied along the Pt stripe, which produce a density current J = 10$^7$ A/cm$^2$. M is the magnetization vector arising through the SHE or OHE. (b) Final Malus curves acquired for S (black circles) and P (pink circles) polarizations (Φ is the goniometer angle of the analyzer, ϕ is the difference between Φ and the angle of the minima of the Malus laws). (c) Analyzer front view scheme showing the polarization axes (polarization and extinction axes) and the positive and negative angles (± ϕ) from extinction (i.e., Φ referenced to the angle corresponding to the minimum of the Malus law). (d) Derivative of the signals reported in panel (b). The perfect linearity of the derivatives (a linear fit gives a R$^2$ = 0.99989) in the ϕ (angle from extinction) range ± 0.8° show that the Malus laws are fully symmetric with respect to extinction, indicating the achievement of a high precision polarization alignment. ± ϕ$_0$ indicate the working angles at which the measurement sensitivity is determined.

Next, measurements are performed at $\pm \phi_0$ ($\pm 0.7°$) from extinction. For a given polarization, the signal recorded by the lock-in at $+\phi_0$ is $Sig_i (+\phi_0) \approx 2\theta_K + f(v_{bias})$ where $\theta_K$ is the Kerr rotation induced by $J(\omega)$ and $f(v_{bias})$ contains the linear $v_{bias}$ contribution to $V^{lock-in}(\omega)$ as well as the fluctuating thermoreflectance term. Similarly, $Sig_i (-\phi_0) \approx -2\theta_K + g(v_{bias})$. Each measurement at a given $v_{bias}$ is performed by changing the analyzer angle from $+\phi_0$ to $-\phi_0$ multiple times (**n**). For each $v_{bias}$, the multiple recorded values of the light intensity transmitted by the analyzer and detected by an avalanche photodiode connected to a lock-in amplifier are averaged leading to our final signals $Sig (\pm\phi_0) = <(Sig_i (\pm\phi_0))_n>$.

Figure 3a show the measured Sig (± φ₀) vs. v_bias for S polarization. As expected, Sig (± φ₀) vs. v_bias are linear with slightly different slopes due to the contribution from residual polarization misalignment (for a perfect alignment, the slope should be identical, namely f(v_bias) = g(v_bias)). Therefore, for S polarization the difference between the two plots ΔSig = Sig (+ φ₀) – Sig (- φ₀) = 4θ_K – (f(v_bias)-g(v_bias)) contains the magneto-optic signal superimposed to a linear v_bias contribution and is displayed in Figs. 2b. The extrapolation ΔSig to v_bias= 0, ΔSig

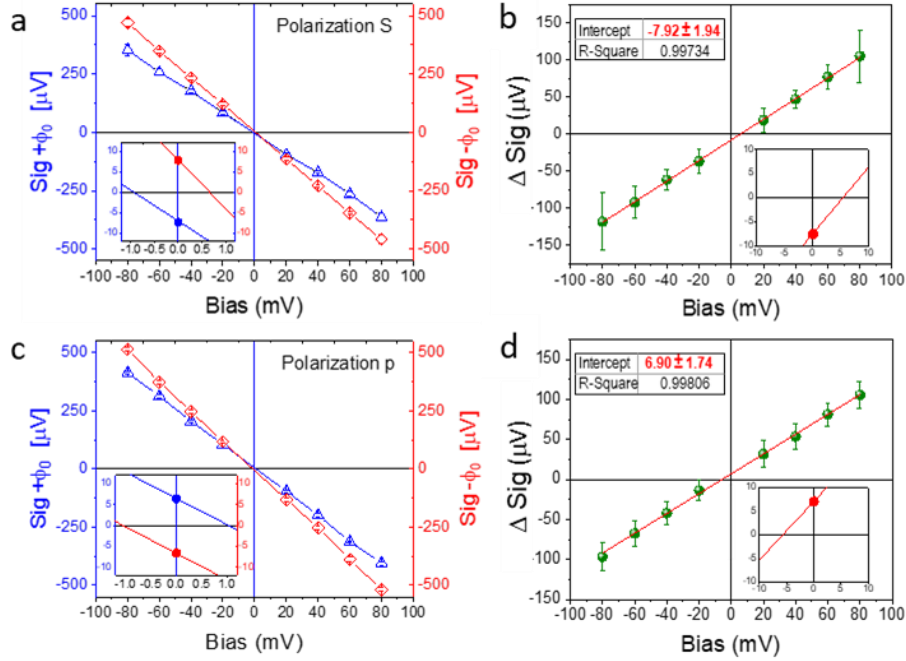

**FIG. 3**. (a) Sig (± φ₀) vs. v_bias signals acquired for positive (blue triangles) and negative (red diamonds) φ₀ values with an incident polarization S. The inset shows a zoomed portion of the plot of the linear fit of the two data sets near the origin. The red and blue dots mark the intercepts of the linear fit lines. (b) Difference of signals Sig (± φ₀), ΔSig, in panel (a) (green spheres). The intercept of the linear fit of the ΔSig data (red line) with the y-axis provides the voltage signal generated solely by the Kerr rotation. In the inset a zoom around the origin of the axes, where the red circle indicates the intercept. Panels (c) and (d) display the same measurements and data treatment of panels (a) and (b), by using the P polarization.

(v_bias→0), gives a value of (-7.92 ± 1.94) µV corresponding to a rotation of $|4\theta_K^S|$. This dual ± φ₀ measurements are equivalent to a pair of measurements by swapping the light path. To further minimize residual symmetry-induced artifacts and to detect and isolate any eventual polar MOKE contributions, the same measurement procedure is repeated using the complementary S polarization state (Fig. 3c and 3d). The extrapolated ΔSig (v_bias→0) for P polarization is equal to (6.90 ± 1.74) µV, corresponding to a rotation of $|4\theta_K^P|$. As expected, the extrapolated ΔSig (v_bias→0) for P and S polarization have the opposite sign (the specific sign depends on the positive direction assumed for the current I, which in our case is that shown in Fig. 2a). By combining these two sets of measurements and after multiplying for √2 to account for the root mean square of the measured voltage provided by the lock-in, overall MOKE-HE signals Δ = | [ΔSig_P(v_bias→0) – ΔSig_S (v_bias→0)]^S | ≈ (11.20 ± 2.74) µV corresponding to $|4\theta_K^S|$, and Δ = | [ΔSig_P(v_bias→0) – ΔSig_S (v_bias→0)]^P | ≈ (9.76 ± 2.46) µV

corresponding to $|4\theta_K^p|$. After applying the calibration, we obtain MOKE-HE rotations $|\theta_K^s| = (7.92 \pm 1.94)$ nrad and $|\theta_K^p| = (6.89 \pm 1.74)$ nrad. We recall here that the reported $|\theta_K^s|$ values for Pt films (~ 50 nm) range from ~ 50 nrad/($10^7$A/cm$^2$) (~ 50 nm films) [12,14] to ~2.5 nrad/($10^7$A/cm$^2$) (20 nm). [16]

It is worth noting that according to the analysis of the optical signal modelled in the Section **7** of the Supporting Information, in case of a perfectly aligned system the intercepts of Sig. + $\phi_0$ and Sig. - $\phi_0$ should be opposite in sign but with the same absolute value (see Fig. S6b), namely |(Sig. + $\phi_0$)$v_{bias}$→0| = |(Sig. - $\phi_0$)$v_{bias}$→0|. This condition is nearly perfectly achieved in our experiment as it shown in the insets of Figs. 3a and 3c, which shows a zoomed portion of the plot of a linear fit of the two data sets Sig. + $\phi_0$ and Sig. - $\phi_0$ near the origin. The red and blue dots mark the intercepts of the linear fit lines, and their position matches excellently the predicted behavior for perfect alignment (*polarization-purity*).

Fig. S6, in particular Panels a and c, shows another relevant result: although the intercepts of measurements in case of tilt are such that |(Sig. + $\phi_0$)$v_{bias}$→0| ≠ |(Sig. - $\phi_0$)$v_{bias}$→0|, the analysis shows that the intercept of the difference (ΔSig) $v_{bias}$→0 = (Sig. + $\phi_0$)$v_{bias}$→0 – (Sig. - $\phi_0$)$v_{bias}$→0 is *not affected* by the tilt. This result, confirmed experimentally, is of the outmost relevance since it would imply that the here proposed methodology relying on determining (ΔSig) $v_{bias}$→0 is self-correcting for small tilts of the incidence plane, i.e., for the presence of a spurious S-polarized (or P-polarized) component for P-polarized (S-polarized) incidence. In order to verify experimentally this crucial property, we conducted an experiment using P-polarized light and inducing a misalignment condition of the polarization vector with respect the optical plane by rotating the polarizer (Polarizer 1) by 0.1° from the optimal angle where P is perfectly parallel to the optical plane. The results of this experiment are reported in Fig. S7. As anticipated by the calculations for this case, the intercepts of the of the two experimental data sets Sig. + $\phi_0$ and Sig. - $\phi_0$ are different and, in particular, |(Sig. + $\phi_0$)$v_{bias}$→0| > |(Sig. - $\phi_0$)$v_{bias}$→0| exactly as predicted (Inset in Fig. S7b). Finally, also in excellent agreement with the modelling prediction, Fig. S7c shows that (ΔSig)$v_{bias}$→0 = (Sig. + $\phi_0$)$v_{bias}$→0 – (Sig. - $\phi_0$)$v_{bias}$→0 return substantially the same value as that obtained for the optimal alignment, thereby confirming experimentally the self-correction mechanism resulting in the effective rejection of the systematic error induced by the tiny spurious polarization component (S-polarized in this case).

A possible last concern comes from the fact that the plots of the signals Sig(+ φ₀) and Sig(- φ₀) vs. $v_{bias}$ in Figs. 3a and 3c (likewise in Fig. S7b) should be perfectly overlapping given that are dominated by the signal $V^{th}(\omega)$ = 2 $R_{s,p}(T)$ $v_{bias}v_0(\omega)$. The non-complete overlapping observed arises from the limited precision and mechanical hysteresis of our automated rotation stage for the Polarizer 2 (analyzer) at setting repeatedly the measurements angles ± φ₀. To check the effect of the uncertainty in the positioning of the rotator, we implemented a slower and more accurate swing between the angles (reducing the angular speed when approaching the targeted angle and approaching it always from the same rotation direction). The results are discussed in Section 9. of the Supporting Information and summarized in Figure S8, and shows that enhancing the rotator repeatability to achieve a closer overlap of the two signals Sig(+ φ₀) and Sig(- φ₀) leads to a substantially unvaried value of the intercept (ΔSig)$v_{bias}$→0, thus a nearly identical value for $θ_K^p$. This result confirms that the polarization alignment method based on the Malus law and the self-correcting mechanism for the effective rejection of the systematic error self-correcting provide the necessary level of measurement accuracy. A final confirmation comes from the measurements carried out using a transverse MOKE configuration as shown in Figure 4, where any current-induced HE is perpendicular to the incidence plane, and thus $θ_K^p$ should vanish producing ΔSig$_P$($v_{bias}$→0) = 0. After a proper alignment for both P and S polarization (see Fig. 4a), the application of our measurement procedure led indeed to nearly vanishing intercept values (ΔSig) $v_{bias}$→0 equal to (-0.24 ±1.98) μV and (0.70 ± 0.86) μV for S and P polarization, respectively, as shown in Figs. 4c and 4d. Within our signal resolution, these results confirm a vanishingly small (< 1 nrad) value of $θ_K$ for this configuration.

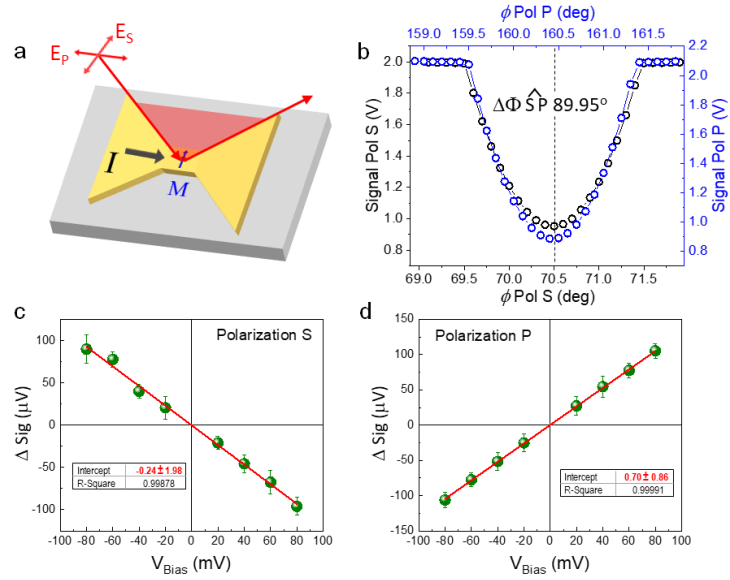

**FIG. 4.** (a) Sample arrangement for optical Kerr rotation experiments for transverse magnetization (b) Malus laws for S and P polarization after optimization of the alignment as described in the main text. Panels (c) and (d) display the plots of ΔSig vs. $v_{bias}$ for S and P polarized light, respectively. The red lines are the linear fit of the data, and the parameters of the fit are reported in the tables.

In conclusion, all the experimental results discussed, corroborated by modelling, confirm that the proposed methodology reliant on a polarization alignment method based on the Malus law, a self-correcting mechanism for the effective rejection of the systematic error, and comparison with measurements with orthogonal incident polarizations (S and P) provides the necessary level of robustness and accuracy to detect MOKE-HE. Furthermore, it is noteworthy that by considering the product of the spin diffusion length and the electrical conductivity of Pt and using the spin Hall conductivity value obtained from ab initio calculations, the resulting values of $\theta_K^i$ (with i either P or S) align reasonably well with that obtained here. [18–20]

**Conclusions**

We have developed a new MOKE-based measuring protocol aiming at a robust and accurate extraction of any current-induced magnetic moment accumulation in metals, such as occurring by charge to spin/orbit conversion, due to spin-orbital Hall effect (MOKE-HE). Having identified the origin of most commons incertitude in conventional MOKE experiments aiming to discern any tiny MOKE-HE signal imbedded in large current-induced thermoreflectance effect and difficult to avoid spurious polarization rotation signals, we have developed a new protocol that targets three crucial steps in measurements: a) light source stability and high sensitive detention, b) robust suppression of any MOKE-HE signal masking by current-induced thermoreflectance, and c) controlled high sensitive method for setting the light polarization and reject leakage of spurious polarization components. Using this approach, which allows faster measurements and self-correcting capabilities for small tilts of the incidence plane and spurious polarization components, we obtained a record sensitivity of 354 nV/nrad and we have obtained specific Kerr rotations of $|\theta_K^S/j| = (7.92 \pm 1.94)$ nrad/($10^7$ A/cm$^2$) for S polarization and $|\theta_K^P/j| = (6.89 \pm 1.74)$ nrad/($10^7$ A/cm$^2$) for P polarization in a 50 nm thick bar of Pt. While the Malus plots (Figs.2b and 4d) allow an accurate determination of the absolute values of Kerr rotation, the error bars in the extracted Kerr rotation angles do not come from averaging measures, as typically done, but from the linear fits of the optical unbalance versus a bias voltage, as illustrated, for instance, in Figs. 3. Both features provide fundamental advances in accuracy and precision compared to conventional MOKE measuring techniques. The extracted value of $|\theta_K^S/j|$ is seven times smaller than previous results on nominally identical Pt devices. [12,14–16] This dramatic difference may come from the measuring methodology, as discussed, or from differences on samples. Indeed, the current-induced Kerr rotation is predicted to depend on a number of parameters characterizing the sample under study. Most noticeable, the density of states at the Fermi level (D($E_F$)), the spin/orbital-Hall conductivity ($\sigma^{S,O}_{xy}$), both with a linear dependence, and the resistivity ($\rho$), with a quadratic dependence. [12] In the resistivity range relevant here, $\sigma^S_{xy}$ of Pt is known to depend only slightly on film resistivity and similarly, [21] the D($E_F$) and the Stoner enhancement factor are not expected to vary significantly for film thicknesses of interest here ((~ 50 nm). In contrast, the quadratic-dependence on resistivity may be relevant. However, as our Pt film, have a resistivity twice as larger than that of Stamm *et al*. [12] films, thus the predicted $|\theta_K^S/j|$ should be much larger (roughly four times) than in Ref. [12]. This is opposite to experimental observation, indicating that difference of

resistivity of Pt films cannot account for the discrepancy of reported $|\theta_K^S/j|$ in our case. Considering that there are also reports of even smaller specific Kerr rotation (i.e., $|\theta_K^S/j| \sim 0.2$ nrad/($10^6$ A/cm$^2$)) for similarly thick Pt structures [12,14–16], our results indicate that experimental procedures and models should be revised accordingly.

**Acknowledgment**

E.L and J.F. acknowledge the financial support of the Spanish Ministry of Science and Innovation through Projects PID2023-152225NB-I00 and Severo Ochoa MATRANS42 (CEX2023-001263-S), supported by MICIU/AEI/10.13039/501100011033 and FEDER, EU. The authors also acknowledge support from Projects TED2021-129857B-I00 and PDC2023-145824-I00, funded by MCIN/AEI/10.13039/501100011033 and the European Union NextGeneration EU/PRTR, as well as from project 2021 SGR 00445 funded by the Generalitat de Catalunya. P.V. acknowledge the Spanish Ministry of Science and Innovation which supported the work at nanoGUNE under the Maria de Maeztu Units of Excellence Program (Grant No. CEX2020-001038-M) and Project No. PID2021-123943NB-I00 (OPTOMETAMAG).

**Contribution**

E.L. and J.F. conceived the experiment. E.L. grew the Pt thin films, performed the X-ray, AFM and electrical characterization, and also did the lithography process. P.V. designed and built the experimental magneto-optical measurement set up and performed the measurements described in the manuscript. All authors discussed the data. E.L. wrote the first draft of the manuscript with inputs from all co-authors. All authors revised the final version of the manuscript.

**Bibliography**

[1]  Z. Q. Qiu and S. D. Bader, Surface magneto-optic Kerr effect, Review of Scientific Instruments **71**, 1243 (2000).

[2]  A. Kimel et al., The 2022 magneto-optics roadmap, J Phys D Appl Phys **55**, 463003 (2022).

[3]  D. Go, D. Jo, C. Kim, and H. W. Lee, Intrinsic Spin and Orbital Hall Effects from Orbital Texture, Phys Rev Lett **121**, 086602 (2018).

[4]  Y. Yao et al., Orbitronics for energy-efficient magnetization switching, Science China Information Sciences **68**, 119402 (2025).

[5]  D. Go and H. W. Lee, Orbital torque: Torque generation by orbital current injection, Phys Rev Res **2**, 013177 (2020).

[6]  J. Sinova, S. O. Valenzuela, J. Wunderlich, C. H. Back, and T. Jungwirth, Spin Hall effects, Rev Mod Phys **87**, 1213 (2015).

[7]  J. Sinova, D. Culcer, Q. Niu, N. A. Sinitsyn, T. Jungwirth, and A. H. MacDonald, Universal intrinsic spin Hall effect, Phys Rev Lett **92**, 1 (2004).

[8]  Y. K. Kato, R. C. Myers, A. C. Gossard, and D. D. Awschalom, Observation of the spin Hall effect in semiconductors, Science **306**, 1910 (2004).

[9]  O. M. J. Van 'T Erve, A. T. Hanbicki, K. M. McCreary, C. H. Li, and B. T. Jonker, Optical detection of spin Hall effect in metals, Appl Phys Lett **104**, 172402 (2014).

[10] Y. Su, H. Wang, J. Li, C. Tian, R. Wu, X. Jin, and Y. R. Shen, Absence of detectable MOKE signals from spin Hall effect in metals, Appl Phys Lett **110**, 42401 (2017).

[11] P. Riego, S. Vélez, J. M. Gomez-Perez, J. A. Arregi, L. E. Hueso, F. Casanova, and A. Berger, Absence of detectable current-induced magneto-optical Kerr effects in Pt, Ta, and W, Appl Phys Lett **109**, 172402 (2016).

[12] C. Stamm, C. Murer, M. Berritta, J. Feng, M. Gabureac, P. M. Oppeneer, and P. Gambardella, Magneto-Optical Detection of the Spin Hall Effect in Pt and W Thin Films, Phys Rev Lett **119**, 087203 (2017).

[13] V. H. Ortiz, S. Coh, and R. B. Wilson, Magneto-optical Kerr spectra of gold induced by spin accumulation, Phys Rev B **106**, 014410 (2022).

[14] Y.-G. Choi, D. Jo, K.-H. Ko, D. Go, K.-H. Kim, H. Gyum Park, C. Kim, B.-C. Min, G.-M. Choi, and H.-W. Lee, Observation of the orbital Hall effect in a light metal Ti, Nature **619**, 52–56 (2023).

[15] I. Lyalin, S. Alikhah, M. Berritta, P. M. Oppeneer, and R. K. Kawakami, Magneto-Optical Detection of the Orbital Hall Effect in Chromium, Phys Rev Lett **131**, 156702 (2023).

[16] Y. Marui, M. Kawaguchi, S. Sumi, H. Awano, K. Nakamura, and M. Hayashi, Spin and orbital Hall currents detected via current-induced magneto-optical Kerr effect in V and Pt, Phys Rev B **108**, 144436 (2023).

[17] D. A. Allwood, G. Xiong, and R. P. Cowburn, Magneto-Optical Kerr Effect Analysis of Magnetic Nanostructures, J. Phys. D: Appl. Phys. **36,** 2175 (2003).

[18] J. C. Rojas-Sánchez, N. Reyren, P. Laczkowski, W. Savero, J. P. Attané, C. Deranlot, M. Jamet, J. M. George, L. Vila, and H. Jaffrès, Spin pumping and inverse spin hall effect in platinum: The essential role of spin-memory loss at metallic interfaces, Phys Rev Lett **112**, 106602 (2014).

[19] M. H. Nguyen, D. C. Ralph, and R. A. Buhrman, Spin Torque Study of the Spin Hall Conductivity and Spin Diffusion Length in Platinum Thin Films with Varying Resistivity, Phys Rev Lett **116**, 126601 (2016).

[20] E. Sagasta, Y. Omori, M. Isasa, M. Gradhand, L. E. Hueso, Y. Niimi, Y. Otani, and F. Casanova, Tuning the spin Hall effect of Pt from the moderately dirty to the superclean regime, Phys Rev B **94**, 060412 (2016).

[21] L. Zhu, L. Zhu, M. Sui, D. C. Ralph, and R. A. Buhrman, Variation of the giant intrinsic spin Hall conductivity of Pt with carrier lifetime, Sci Adv **5**, eaav8025 (2019).

# High accuracy Spin Hall Effect Induced Spin Accumulation detection in MOKE Measurements

Emanuele Longo[1,*], Josep Fontcuberta[1,] Paolo Vavassori [2,3,**]

1. Institut de Ciència de Materials de Barcelona (ICMAB-CSIC), Campus UAB, Bellaterra, Catalonia 08193, Spain
2. CIC nanoGUNE BRTA, E-20018 Donostia-San Sebastian, Spain
3. IKERBASQUE, Basque Foundation for Science, E-48009 Bilbao, Spain

*elongo@icmab.es, ** p.vavassori@nanogune.eu

*Keywords: Magneto Optic, Spin Hal effect, Kerr rotation Platinum*

**Supporting Information**

1. **X-ray characterization**

Figure S1 shows the chemical-structural characterization of a nominally 50 nm thick sputtered Pt thin film. In panel (a) the grazing incidence X-ray diffraction (GIXRD) pattern acquired for a rocking angle $\omega = 1.2°$ is reported. The (111) reflection placed around $2\theta = 39.6°$ is the main peak of GIXRD peak pattern, while a less intense (200) appear at $2\theta = 46.0°$, features in agreement with the powder diffraction pattern of the cubic crystal structure of Pt.[1] To investigate on the Pt texturization, the XRD pattern acquired in Bragg-Brentanto geometry is reported. A clear and sharp peak positioned at $2\theta = 39.6°$ is present, and it is attributed to the (111) crystalline reflection. The thickness, roughness and electronic density of the Pt films are extracted from the fit of the X-ray reflectivity data shown in Figure S1d and the values are reported in Table S1. The Pt electronic density ($\rho_e$) is 4.957 $e^-/Å^3$, a value 5% lower than the nominal one (5.191$e^-/Å^3$), in accordance with the thin film nature of the material.

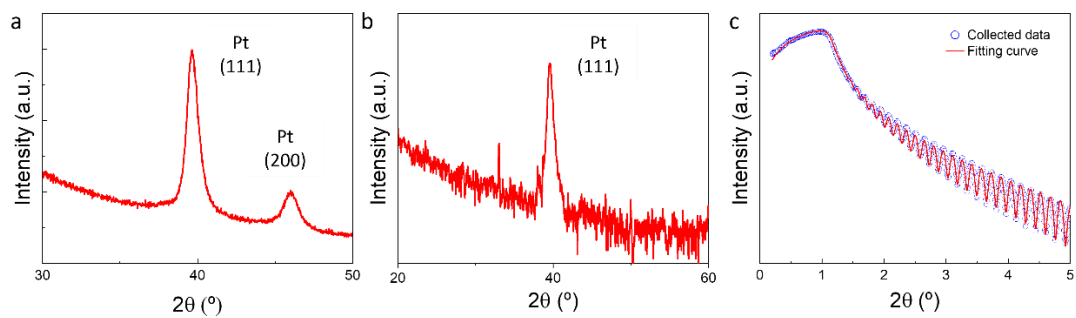

**Fig. S1.** X-ray characterization of the Pt film. (a) and (b) show the grazing incidence X-ray diffraction pattern and the Bragg-Brentano scan, respectively. The number in brackets indicate the Miller indices of the corresponding reflections of the cubic Pt phase. (c) X-ray reflectivity collected spectra (blue circles) and corresponding fit performed with Croce-Nevot model (red solid line). The parameters extracted from the fit are reported in Table S1.

**TAB. S1.** Parameters extracted from the fit of the XRR curve reported in Figure S1 (c).

|  | Thickness (nm) | $\rho_e$ ($\frac{e^-}{Å^3}$) | Roughness (nm) |
|---|---|---|---|
| Pt | 53.3 | 4.957 | 0.23 |
| SiO$_2$ | / | 0.674 | 0.21 |

2. **Surface morphology and step profile of the lithographed Pt stripes**

Figure S2 (a) and (b) displays the atomic force microscopy (AFM) images of a Pt stripe surface of two different areas of 5x5 µm² and 1x1 µm², respectively. The morphological analysis of the latter images indicates a surface roughness of 0.313 nm and 0.297 nm for panel (a) and (b), respectively. These values are in accordance with the XRR analysis as detailed in Figure S1. Panels (c) shows an AFM image of 15x15 µm² acquired on one edge of a Pt stripe. The black dashed line indicates the direction of the height profile reported in panel (d) of the same figure, from which a step of 57 nm is measured. The thickness measured with AFM profilometry is in accordance with the one measured with XRR, as reported in Table S1.

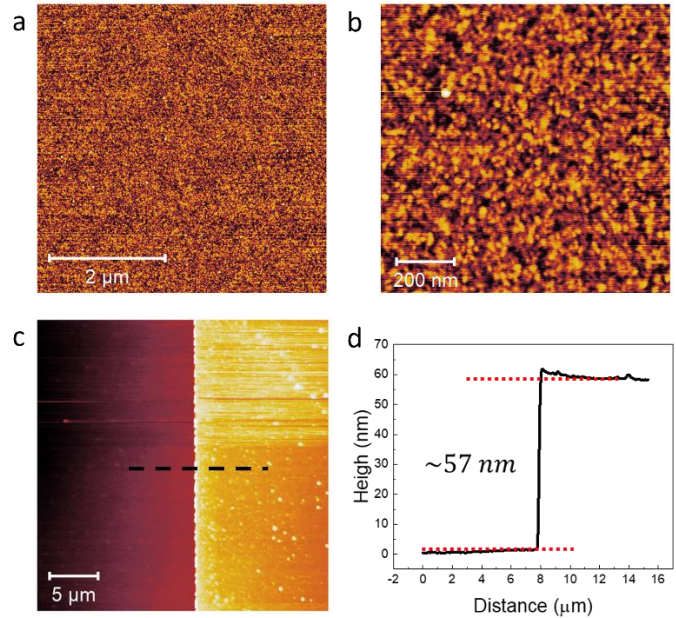

**Fig. S2.** AFM analysis of the Pt thin film surface. (a) and (b) represent the AFM images of two areas of 5x5 µm² and 1x1 µm², respectively. Panel (c) is an AFM image along the edge of a patterned Pt stripe. In panel (d) is reported the heigh profile of the Pt stripe edge, as extracted along the black dashed line in panel (c).

### 3. Transport measurement in Pt stripes

Figure S3 represents the resistivity ($\rho_0$) of a Pt stripe as a function of the temperature acquired with a Quantum Design PPMS setup in a crossbar configuration. From a linear fit we extracted a slope of $(7.21 \pm 0.01) \cdot 10^{-4}\ \mu\Omega \cdot cm/K$. The $\rho_0$ value we extracted at room temperature for the Pt thin film studied in the manuscript is about two times larger than the one

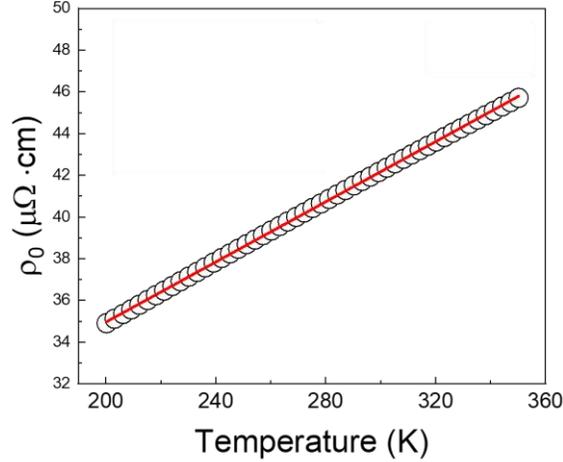

**Fig. S3.** Resistivity measurement as a function of temperature, acquired in four-point configuration.

reported in Ref.[2].

### 4. Magnitude of thermoreflectance in MOKE experiment

Considering $R^{2\omega}$ as the signal of the APD measured from the lock-in, in our experiments we have a $R^{2\omega}$ of approximately -3.5 mV, that compare well with the values of -0.232 mV provided by Ref.[2] authors in a private communication, given the ten times higher sensitivity of our system.

Our sample is slightly more resistive than the one used in Ref. STAMM, thus the thermoreflectance is higher in our samples $\frac{\Delta R}{R} = -1.5 \cdot 10^{-3}$, while in Figure 2 of Ref.[2] the authors reported a value of about $-0.9 \cdot 10^{-3}$. This makes the two experimental condition comparable, although in our case the temperature is approximatively 10ºC higher (assuming a thermoreflectance factor of $0.5 \cdot 10^{-4} \frac{\Delta R}{\Delta T}$ in the visible range).[3,4]

5. **Experimental setup**

Figure S4 represent a schematic of the experimental setup.

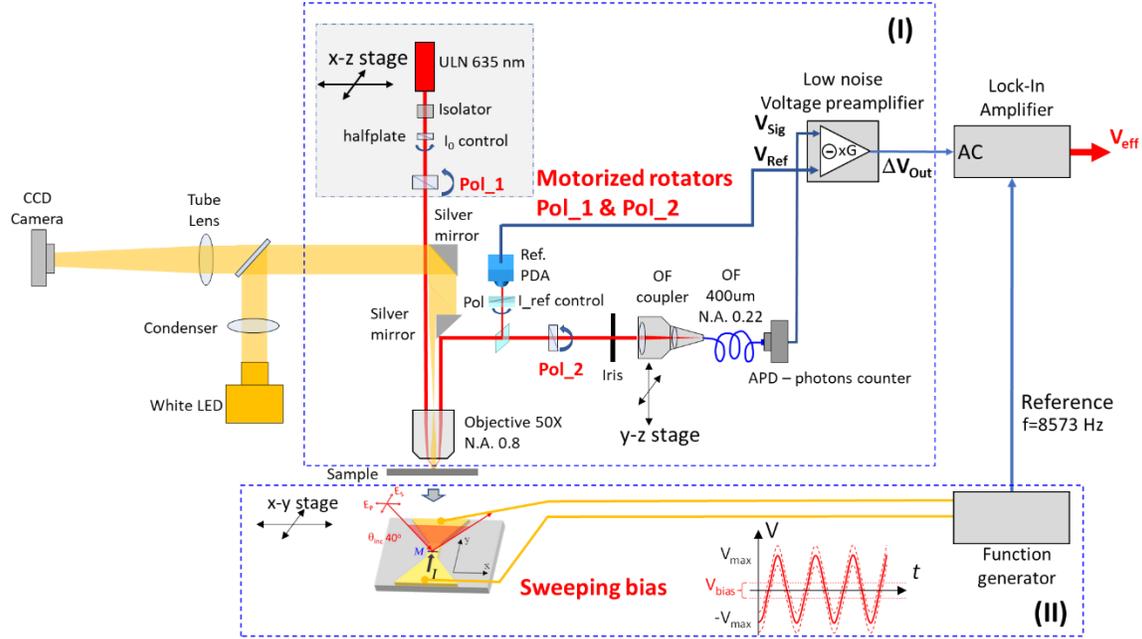

**FIG. S4.** Scheme of the magneto-optical experimental setup.
The different components are:
- Laser source: Ultra-low noise ULN 635 Coherent; power 5 mW, wavelength 635nm linearly polarized, RMS noise <0.06% for bandwidths of 10 Hz to 10 MHz; https://www.coherent.com/lasers/diode-modules/ultra-low-noise.
- APD: APD440A2, UV enhanced Si avalanche photodiode (peak responsivity @ 600nm); NEP 2.5 fW/√Hz BW 100 kHz; Thorlabs: https://www.thorlabs.com/thorproduct.cfm?partnumber=APD440A2.
- PDA: Amplified photodiode Thorlabs PDA36A; Gain 30 dB NEP= 3.4 pW/ √Hz BW 260 kHz; Thorlabs: https://www.thorlabs.com/thorproduct.cfm?partnumber=PDA36A2.
- Polarizers: SM1PM10 Glan-Taylor polarizing prisms, Extinction ratio 10^5:1; Thorlabs: https://www.thorlabs.com/thorproduct.cfm?partnumber=SM1PM10#ad-image-0;
- Motorized rotators: Model SR50 CC bidirectional accuracy 0.015degs = 260 urad; Newport: https://www.newport.com/p/SR50CC.
- Half plate: Zero order half plate WPH10M-633; Thorlabs: https://www.thorlabs.com/thorproduct.cfm?partnumber=WPH10M-633.
- Objective: Olympus MPLFN50X Objective Polarization maintaining 50X 0.8 NA; https://www.edmundoptics.eu/p/olympus-mplfn50x-objective/55560/.
- Function generator: TG5011 frequency resolution 1μHz, harmonic distortion -65dBc (DC to 20kHz), phase noise -115dBc/Hz; https://www.aimtti.com/product-category/function-generators/aim-tg251xa-501xaseries.
- Low noise differential amplifier: Signal Recovery 5113 PRE-AMP low noise voltage preamplifier with continuously adjustable gain and selectable high, low and bandpass filtering; https://www.ameteksi.com/products/lock-in-amplifier-en/5113a-low-noise-preamplifier.
- Lock-In: Signal Recovery 7265 Dual Phase DSP Lock-in Amplifier; Voltage noise 2 nV/√Hz @ 1 kHz, C.M.M.R. > 100 dB @ 1 kHz, frequency response 0.001 Hz to 250 kHz; https://www.ameteksi.com/support-center/legacy-products/signal-recovery-legacy/lock-in-amplifier-legacy/7280-wide-bandwidth-dsp-lock-in-amplifier .

The sample is mounted on a piezo xy-stage to allows placing the focused laser beam precisely in the desired position on the Pt microstructure (in our case away from the edges of the Pt channel. For the measurements we used an ultra-low noise red laser at 635 nm wavelength and attenuated to a power of 1 mW using a linear polarizer (the laser is linearly polarized at 45°, i.e., along the bisector of x and y axes) placed before the polarizer Pol_1, which defines the final incident polarization. The polarize Pol_1 is mounded in a motorized rotational stage. The linearly polarized beam travels unperturbed until a 50x polarization maintaining objective of high numerical aperture NA = 0.8, used to focus the beam on the sample at a angle of incidence of 40° (achieved by an horizontal displacement of the laser beam from the objective axis). The laser is forming an elliptical spot (~ 4 μm x 5 μm) on the sample surface. The reflected beam from the sample surface is collected by the objective and sent to the polarization sensitive measurement stage by an Au mirror. The intensity of the reflected beam is monitored by deflecting with a beam splitter part of the reflected beam (10%), which is sent to an amplified photodetector through a polarized for a fine adjustment of the reference voltage $V_{Ref}$. The beam transmitted by the beam splitter goes through the analyzer, Pol_2, which is mounded in a motorized rotational stage. The transmitted beam goes through an iris to block stray light and on to an avalanche Si photodiode (APD, nearly single photon detector) through a collection optics mounted on a *xy* mechanical stage and coupled to the APD sensor (1 mm diameter) via an optical fiber (fiber core 200 μm) to further reduce any remnant stray light. The signal of the APD, $V_{Sig}$, is subtracted to $V_{Ref}$ by low noise differential amplifier set to a gain G = 3 (selected after evaluating the best S/N ratio). The output of the differential amplifier goes to a lock-in amplifier referenced to the frequency of 8.573 kHz of the voltage-controlled current source used to generate the current I(ω) across the Pt bar as well as to apply a variable voltage bias $v_{bias}$. The time constant of the lock-in is set to T = 2 s; before reading via computer (GPIB interfaced) the lock-in output, we wait for 6*T = 12 s for the signal to stabilize and then voltage output of the lock-in is read at the rate of 1sample/ms for 10*T = 20 s. The recorded stream of voltages is then averaged giving the final value $V_{eff}$. For each bias voltage, $V^i_{eff}$ (+$\phi_0$) and $V^i_{eff}$ (-$\phi_0$) are recorded, where ±$\phi_0$ are the selected analyzer angles (see main text). This process is repeated *n* = 16 times for each $v_{bias}$ and the sequence of values for each $\phi_0$ are averaged leading to the final quantities $Sig_{\pm\phi0}$ = <$V^i_{eff}$ (±$\phi_0$) >$_n$ (see Figures 2, S8, and S9). This corresponds to approximately 20 minutes for each bias voltage. Considering a typical measurements were 8 values of $v_{bias}$ are used, the total time taken by a full measurement at a given polarization is of nearly 2.5 hours.

### 6. DC bias approach and effect on the injected I (Omega) current

In Figure S5(a) a sketch of the experimental configuration we adopted is shown. The polarized light (i.e., P - parallel or S -perpendicular to the incidence plane) is directed towards the sample surface from the polarizer with a fixed incidence angle $\theta_{inc} = 40°$ and subsequently detected by analyzer positioned at the specular angle of the polarizer within the incidence plane. The black arrow indicates the injected charge current (I($\omega$)), which generates an accumulation of magnetic moments, with a direction (M) perpendicular to I($\omega$), and in this case parallel to the incidence plane (Fig. S5(a)). Panel (b) of Figure S5 represents the ideal case of a MOKE experiment, where two main signal components are present in the signal $V^D(\omega, 2\omega)$ generated by the APD. $V^K(\omega)$ is the voltage signal of frequency $\omega$ which shall include any current induced magneto-optic response of the system (i.e., Kerr rotation); while $V^{th}(2\omega) \propto I^2$ is the thermo-reflectance component of frequency $2\omega$. (see main text for the details). In a real experiment, $V^K(\omega) \propto 10^{-6}$ and $V^{th}(2\omega) \propto 10^{-3}$, thus being these signal orders of magnitude different. $V^{th}(2\omega)$ is also subjected to significant fluctuation, due to the intrinsic thermal noise and external factors, like tiny temperature variation of the laboratory environment.

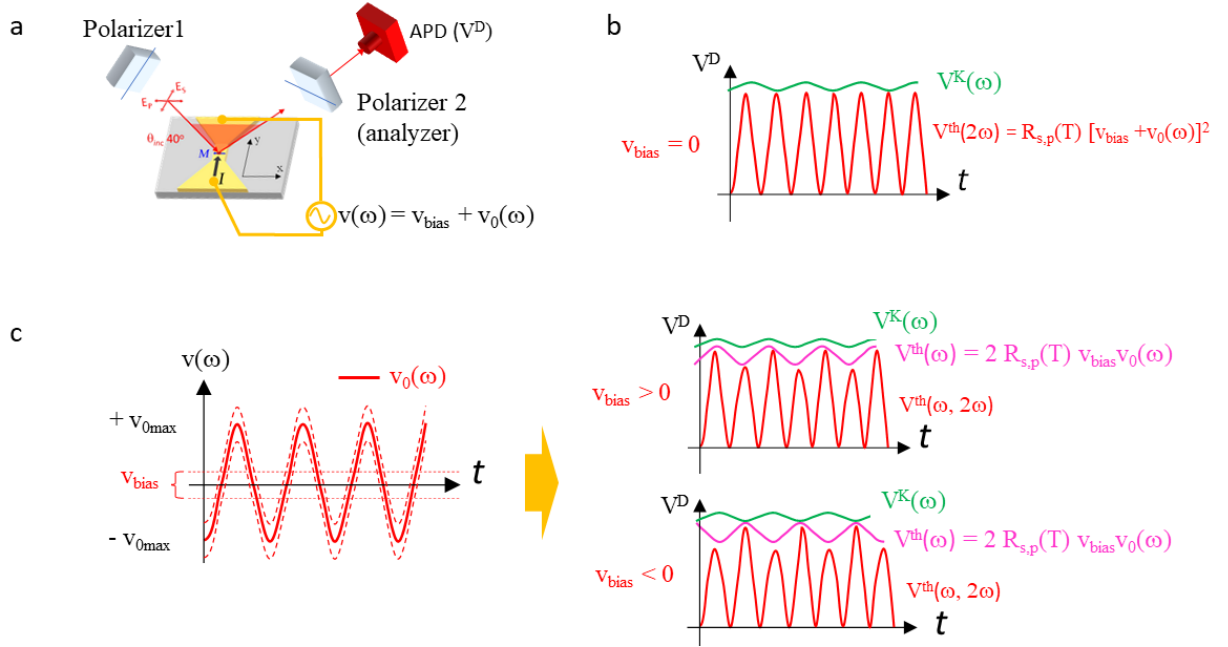

**Fig. S5**. (a) Sketch of the MOKE configuration scheme. The Polarizer1 sets the incident light polarization with an electric field E oscillating along the direction of the red double arrow. The reflected light is collected by the Polarizer 2 (analyzer) while a current I is swept sinusoidally at frequency $\omega$ by a voltage v($\omega$) applied at the Pt stripe. M is the magnetization vector arising through the SHE. (b) MOKE signal (green solid line) and spurious thermal signal (red solid line) for $v_{bias}$= 0. (c) Example of the application of a positive/negative DC voltage bias (red solid line). The magenta curve $V^{th}(\omega)$ indicates the new oscillating voltage resulting by the application of $v_{bias}$.

Such variations produce extra noise in $V^D(\omega)$ term filtered by the lock-in amplifier, further complicating the detection of the already very small magneto-optic response ($\propto 10^{-6}$).

Furthermore, any spurious contribution accompanying the voltage generated by the source may lead to a non-quadratic response in the thermal component $V^{th}(2\omega)$. As a result, undesired linear contributions $V^{th}(\omega)$ cannot be effectively filtered through lock-in amplification, aspect that constitutes a subtle drawback for a correct signal detection. The proposed novel strategy to overcome the aforementioned limitations is based on the introduction of a controlled DC voltage bias $v_{bias}$ superimposed onto the frequency-dependent signal (i.e.,

I(ω)) injected into the thin film structure (the Pt film). This approach serves a two-fold purpose. First, the Lock-in signal at $\omega$ exhibits significant instability, even when employing long time constants at the Lock-in amplifier (e.g., 10 - 20 s, compared to 2 s typically used in the literature). Despite this extended integration time, the signal remained highly fluctuating, with variations exceeding the magnitude expected for a signal corresponding to a Kerr rotation of ~50 nrad (as reported in prior studies Ref. [2]), effect primarily attributed to the limited stability of the thermoreflectance signal. So, instead of averaging over hundreds of repeated measurements, which requires extensive acquisition times and still yields high variance, we found that introducing a DC voltage bias and measuring the response at several $v_{bias}$ values (e.g., 8 values symmetrically spanning positive and negative voltages) significantly reduces the uncertainty. In Figure S5c, a schematic illustrates the influence of applying a bias voltage with $v_{bias} > 0$ and $v_{bias} < 0$, respectively. When a bias different from zero is applied, the system experiences asymmetric heating affecting the thermoreflectance response: the sample is heated more or less during the positive half-cycle of the voltage with respect to the negative half-cycle. Figure S5(c) highlights that this asymmetry introduced by $v_{bias} > 0$ results in a modulation of the thermoreflectance signal $V^{th}(\omega)$ and the amplitude of this modulation is linearly dependent on the magnitude of the applied bias. Conversely, applying a negative bias ($v_{bias} < 0$) introduces a 180º phase shift in $V^{th}(\omega)$, thus providing a response with opposite sign, as shown in Figure S5c. The applied $v_{bias}$ linearly shifts part of the thermal response component at $2\omega$ into the $\omega$ channel. Since the resulting bias-induced signal at $\omega$ is linearly dependent on $v_{bias}$, we can extract the value at zero bias by fitting the data to a linear function and evaluating the intercept, as discussed in the following. Noteworthy, this method leads to a more robust determination (lower noise) of $V^K(\omega)$ given that $V^{th}(\omega)$ is phase locked to $V^K(\omega)$, as confirmed by our experimental data. The second benefit of using the bias technique resides in the mitigation of systematic errors arising from small, unintentional DC offsets produced by the function generator or circuit. Such offsets will invert sign upon swapping the current injection leads, a standard practice used to verify magnetic origin by checking signal inversion. However, even a small residual bias can produce a signal at $\omega$ that mimics the expected symmetry of a magnetic response (e.g., spin or orbital accumulation), thereby introducing false positives. By scanning the bias and extracting the zero-bias intercept, we effectively suppress also this artifact. In summary, bias scanning provides both improved signal-to-noise performance and eliminates systematic errors that may otherwise be misinterpreted as genuine spin or orbital accumulation effects upon reversal of current polarity.

7. **Effect of a tilted plane of incidence (i.e., spurious polarization)**

In case of perfect alignment (no tilted incidence plane) and incident polarization P (the same analysis can be easily carried out for S polarization leading to identical results), the reflected field, can be calculated as follows:

$$\begin{bmatrix} r_{pp} & r_{ps} \\ r_{sp} & r_{ss} \end{bmatrix} \begin{bmatrix} 1 \\ 0 \end{bmatrix} = \begin{bmatrix} r_{pp} \\ r_{sp} \end{bmatrix} = r_{pp} \begin{bmatrix} 1 \\ \frac{r_{sp}}{r_{pp}} \end{bmatrix} = r_{pp} \begin{bmatrix} 1 \\ \theta_K + i\varepsilon_K \end{bmatrix} \text{ given that } \frac{r_{sp}}{r_{pp}} = \widetilde{\Theta}_K = \theta_K + i\varepsilon_K, \text{ namely the complex Kerr}$$

angle. Normalizing to the reflectivity $r_{pp}$, the reflected field can be written as follows $\begin{bmatrix} 1 \\ \widetilde{\Theta}_K \end{bmatrix}$.

The normalized intensity detected by the APD, $\frac{I_{APD}}{I_0}$, can be calculated as follows:

$$\frac{I_{APD}}{I_0} \propto \left\| \begin{bmatrix} \sin^2\phi & \sin\phi\cos\phi \\ \sin\phi\cos\phi & \cos^2\phi \end{bmatrix} \begin{bmatrix} 1 \\ \widetilde{\Theta}_K \end{bmatrix} \right\|^2, \text{ where } I_0 \text{ is } |r_{pp}|^2.$$

The calculation gives a term $\frac{I}{I_0}$ independent of $\widetilde{\Theta}_K$:

$$\frac{I}{I_0} = \sin^2\phi + i_d,$$

where the term $i_d = 1.6 \times 10^{-4}$ was added to account for the experimental depolarization, and the MOKE term:

$$\frac{\text{Delta\_I}}{I_0} = \sin2\phi \cdot \theta_K + O^2(\theta_K, \varepsilon_K),$$

which linearly depends only on the Kerr rotation $\theta_K$ in the first order (the higher order term $O^2(\theta_K, \varepsilon_K)$ is negligibly small).

In case of a small S-polarized component arising from a minute tilt of the reflection plane ($\pm 0.1°$, which in an experiment can be due to a vertical tilt of the incident beam of $\pm 0.1°$ that can be originated by a variety of factors including the beam propagation across optical devices like thick polarizing prisms, the objective, or even to a slight tilt of the sample), the incident field can be written as $\begin{bmatrix} 1 \\ \delta \end{bmatrix}$ considering $\delta = \sin(\pm 0.1°) \ll 1$ consistently with case discussed here, and the reflected field normalized to $r_{pp}$ can be easily derived as follows $\begin{bmatrix} 1 - \delta\widetilde{\Theta}_K \\ \widetilde{\Theta}_K + \delta\widetilde{\alpha} \end{bmatrix}$, where we have considered that $\frac{r_{ps}}{r_{pp}} = \frac{-r_{sp}}{r_{pp}} = -(\theta_K + i\varepsilon_K) = -\widetilde{\Theta}_K$ and $\widetilde{\alpha} = r_{ss}/r_{pp} = \alpha + i\beta$, which for an incidence angle of 40° results in $\widetilde{\alpha} = -0.997972 + i0.1744$ at 635 nm wavelength for Pt.

In this case, the calculation of $\frac{I_{APD}}{I_0}$ is as follows:

$$\frac{I_{APD}}{I_0} \propto \left\| \begin{bmatrix} \sin^2\phi & \sin\phi\cos\phi \\ \sin\phi\cos\phi & \cos^2\phi \end{bmatrix} \begin{bmatrix} 1 - \delta\widetilde{\Theta}_K \\ \widetilde{\Theta}_K + \delta\widetilde{\alpha} \end{bmatrix} \right\|^2.$$

Now, the calculation, adding depolarization, gives a Malus law:

$$\frac{I}{I_0} = \sin^2\phi + \delta\alpha\sin2\phi + i_d + O^2(\delta\alpha)$$

and a MOKE term

$$\frac{\text{Delta\_I}}{I_0} = \sin2\phi \cdot \theta_K + 2\delta(\alpha\cos^2\phi - \sin^2\phi) \cdot \theta_K + O^2(\theta_K, \varepsilon_K, \delta, \alpha, \beta).$$

The calculated sensitivity, neglecting the negligibly small 2nd order terms in $\theta_K$ and $\varepsilon_K$, is thus:

$$S(\phi) = \frac{\text{Delta\_I}}{\theta_K} = \sin2\phi + 2\delta(\alpha\cos^2\phi - \sin^2\phi) \approx \sin2\phi - 2\sin(\pm 0.1°);$$

considering that for the case of interest here $\alpha \approx -1$ (the term in red is that added by the tilt).

Fig. S6 displays the plot of the calculated MOKE sensitivity $S(\phi)$. As shown in Fig. S6b for the case of perfect alignment at $\pm\phi_0 = \pm 0.7°$ the calculations provide sensitivity $S(\pm\phi_0) = 355$ nV/nrad in agreement with the 354 ($\pm 27$) nv/nrad determined experimentally (see main text).

Figs. S6a and S6c show the Malus law and the MOKE sensitivity $S(\phi)$ for a tilt of +0.1° and -0.1°, respectively. The MOKE sensitivity plots in Fig. S6 indicate that a tilted reflection plane results in a $S(\phi)$, and thus a MOKE signal, that is different for $+\phi_0$ and $-\phi_0$ (Figs. S6a and S6c), Therefore the intercepts in our measurements should show that $|(Sig. +\phi_0)v_{bias}\rightarrow 0| \neq |(Sig. -\phi_0)v_{bias}\rightarrow 0|$, while for the perfectly aligned system $|(Sig. +\phi_0)v_{bias}\rightarrow 0| = |(Sig. -\phi_0)v_{bias}\rightarrow 0|$. Nonetheless, the $S(\phi)$ plots indicate that the intercept of difference $(\Delta Sig)\ v_{bias}\rightarrow 0 = (Sig. +\phi_0)v_{bias}\rightarrow 0 - (Sig. -\phi_0)v_{bias}\rightarrow 0$ in our experiments should not be affected by the tilt, given that $(S(+\phi_0) - S(-\phi_0))$ is not affected by the tilt. Therefore, the here proposed methodology relying on $(\Delta Sig)\ v_{bias}\rightarrow 0$ is self-correcting for small tilts of the incidence plane, i.e., for the presence of a spurious S-polarized (or P-polarized for S-polarized incidence) component. This is the key advantage of the here proposed measurement setup and methodology, which would be rather complicated to implement in other detection systems based on balanced detection.

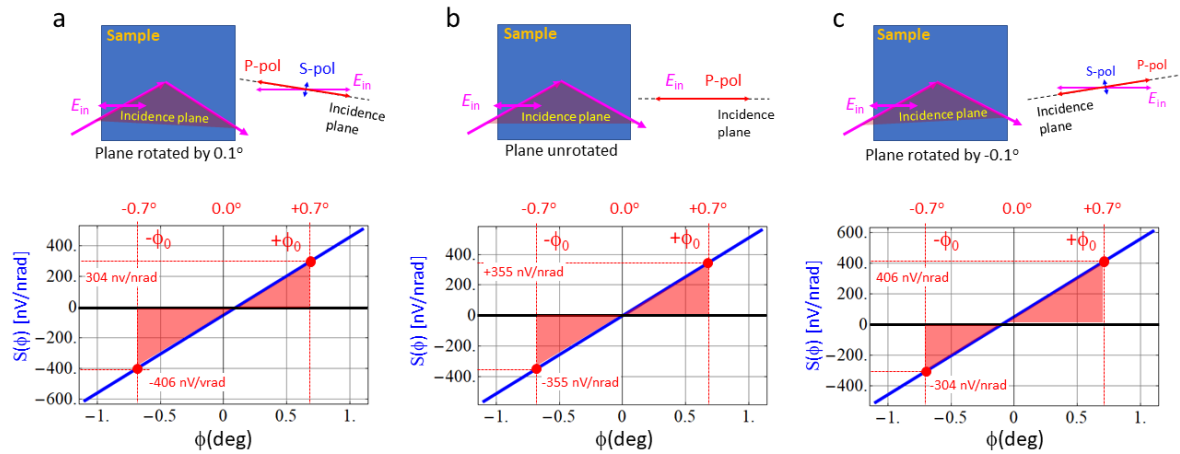

**Figure S6** MOKE sensitivity $S(\phi)$ (see text for the detailed calculations and definition of quantities), calculated for a tilt of $\pm 0.1°$ of the polarization of the incident beam, assumed to be initially P-polarized, equivalent to a tilt in the incidence plane of $\pm 0.1°$ (e.g., caused by a vertical tilt of $\pm 0.1°$ of the incident beam). Panel **a** show the case of a +0.1° tilt; panel **b** is the case of the perfectly aligned polarization; panel **c** shows the case of a -0.1° tilt. In all plots, a depolarization factor of $1.6 \times 10^{-4}$ is assumed. The red labels in the bottom panels mark the values of the MOKE sensitivity at the angles $\pm\phi_0$ equal to $\pm 0.7°$.

8. **Experimental verification of misaligning the polarization vectors with respect to the optical plane**

In order to check experimentally the predictions of the calculations described in the previous section **7**, we conducted an experiment by inducing a misalignment condition of the polarization vector P with respect the optical plane, as shown in Figure S7. The polarizer (Polarizer 1) is rotated 0.1° from the optimal angle where P is perfectly parallel to the optical plane. In panel (a) the Malus curve is shown in this condition (orange points) and compared with one obtained for a "pure" P polarization (magenta symbols). As expected, the minimum of the orange Malus plot is shifted by +0.1°. Panel (b) shows the lock-in signals Sig($+\phi_0$) and Sig($-\phi_0$) vs $v_{bias}$ acquired in this geometry and the inset shows a zoomed in portion of the plot where the continuous lines are linear fits of the two plots. The blue and red dots in the inset mark the intercepts of the two linear fits, namely (Sig. $+\phi_0$)$v_{bias} \rightarrow 0$ and (Sig. $-\phi_0$)$v_{bias} \rightarrow 0$, respectively. As predicted by the calculations of S($\phi$) and $\Delta$S($\phi$) (see Fig. S6 and S7, respectively), for this case $|S(+\phi_0)| < |S(-\phi_0)|$ and $|(Sig. +\phi_0)v_{bias} \rightarrow 0| < |(Sig. -\phi_0)v_{bias} \rightarrow 0|$ in excellent agreement with the experimental plots of Sig($+\phi_0$) and Sig($-\phi_0$) vs. $v_{bias}$ (Fig. S7b) as well as with the different intercept values (see arrows in the inset of Fig. S7). Finally, also in excellent agreement with the modelling prediction, panel (c) shows that ($\Delta$Sig) $v_{bias} \rightarrow 0$ = (Sig. $+\phi_0$)$v_{bias} \rightarrow 0$ - (Sig. $-\phi_0$)$v_{bias} \rightarrow 0$ remained substantially unvaried (7.78 ±2.01 µV) with respect to that obtained for the optimal alignment (6.9 ±1.4 µV for P polarization and -7.9 ±1.94 µV for S polarization), thereby confirming experimentally the self-correction mechanism resulting in the effective rejection of the systematic error (a bias) induced by the tiny spurious polarization component (S-polarized in this case).

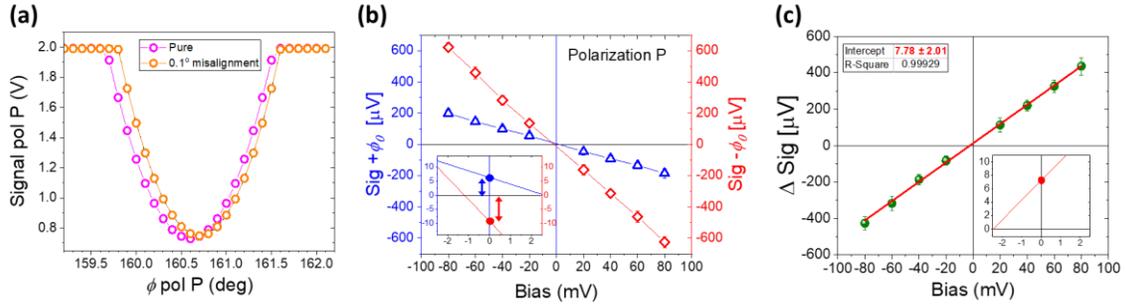

**Fig. S7.** Experimental verification of the effect of a tiny tilt of the incident polarization on the measurements. (a) The Polarizer 1 is rotated by $\Delta\Phi = +0.1°$ from the optimal angle for pure P polarization, then a Malus curve is acquired (orange symbols) and displayed together with that for the optimal configuration (magenta symbols). (b) Signals Sig($+\phi_0$) and Sig($-\phi_0$) vs $v_{bias}$ acquired after the induced polarization tilt $\Delta\Phi = +0.1°$. The inset shows a close view of the intercepts (blue and red dots) of lines obtained by the linear fit of the two signals. Panel (c) displays the difference plot of the two signals, $\Delta$Sig, and its linear fit (red line and values of the fit parameters in the table), together with the zoomed view of the intersection point ($\Delta$Sig) $v_{bias} \rightarrow 0$.

## 9. Sensitivity on the rotators positioning

The plots of the Lock-in signals Sig(+$\phi_0$) and Sig(-$\phi_0$) vs $v_{bias}$ should be perfectly overlapping given that are dominated by the signal $V^{th}(\omega) = 2\, R_{s,p}(T)\, v_{bias} v_0(\omega)$ (see section 6. above). As demonstrated in the previous section 8. for the case of 0.1° misalignment, the non-complete overlapping observed in Figs. 3a, 3c, indicates that there must be a slight misalignment also in those cases. The question naturally arises of when a misalignment can be considered negligible. We implemented a slower and more accurate swing between the angles (reducing the angular speed when approaching the targeted angle and approaching it always from the same rotation direction). The results shown in Figure S8, demonstrates that enhancing the alignment and thereby improving the polarization purity results in the near-complete overlap of the two corresponding plots (see panel (a)). The intercept extracted in Fig. S8(b), ($\Delta$Sig) $v_{bias}\rightarrow 0$ (7.19 ±1.76 µV, returns a substantially unvaried value with respect to that reported in Figure 3 of the main text, as expected given the self-correcting properties of our measurements protocol. The comparison between the insets of Fig. 3a and 3c with the inset of Fig. S8a, shows that the intercepts of the individual signals Sig(+$\phi_0$) and Sig(-$\phi_0$) remain substantially unvaried, namely |(Sig. +$\phi_0$)$v_{bias}\rightarrow 0$| = |(Sig. -$\phi_0$)$v_{bias}\rightarrow 0$|, indicating that the alignment achieved based on the recursive Malus laws protocol described in the text is already sufficient to achieve a sufficiently high polarization purity, without the need of a much more time consuming approach.

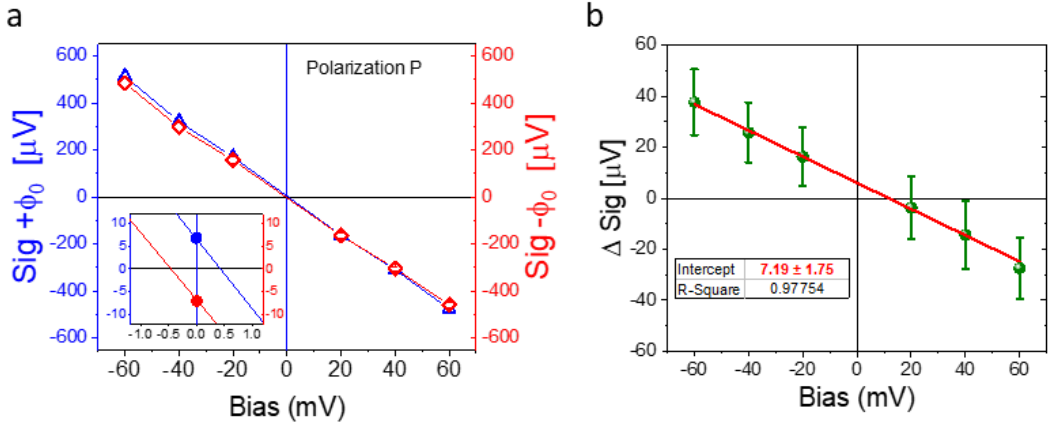

**Fig. S8.** (a) Lock-in signal acquired for two positions of the polarizer ± $\phi_0$ using a more accurate displacement of the rotator of Polarizer 2. (b) Difference of the two lines in panel (a) (green spheres). From the linear fit (red solid line) the intercept value is extracted and compared with those reported in Figs. 3c and S7c.

**Bibliography**

1. Longo, E. *et al.* ALD growth of ultra-thin Co layers on the topological insulator Sb2Te3. *Nano Res* **13**, 570–575 (2020).
2. Stamm, C. *et al.* Magneto-Optical Detection of the Spin Hall Effect in Pt and W Thin Films. *Phys Rev Lett* **119**, 087203 (2017).
3. Favaloro, T., Bahk, J. H. & Shakouri, A. Characterization of the temperature dependence of the thermoreflectance coefficient for conductive thin films. *Rev. Sci. Instrum.* **86**, 024903 (2015)
4. Wilson, R. B. *et al.* Thermoreflectance of metal transducers for optical pump-probe studies of thermal properties. *Optics Express* **20**, 28829 (2012).